\documentclass[preprint,aps,showkeys,showpacs,preprintnumbers,amsmath,amssymb]{revtex4}

\usepackage{graphicx}
\usepackage{dcolumn}
\usepackage{bm}

\begin{document}


\title{Relaxation of thermo-remanent magnetization in Fe-Cr GMR multilayers}

\author{R. S. Patel}
\author{A. K. Majumdar}
 \email{akm@iitk.ac.in}
\affiliation{Department of Physics, Indian Institute of Technology,
Kanpur-208 016, India}

\author{A. K. Nigam}
\affiliation{Tata Institute of Fundamental Research, Homi Bhabha Road,
Mumbai-400 005, India}

\date{\today}

\begin{abstract}

The time decay of the thermo-remanent magnetization (TRM) in Fe-Cr
giant magnetoresistive (GMR) multilayers has been investigated.
The magnetization in these multilayers relaxes as a function of time after
being cooled in a small magnetic field of 100 Oe to a low temperature
and then the magnetic field is switched off. Low-field
($<$ 500 Oe) magnetization studies of these samples have shown
hysteresis. This
spin-glass-like behavior may originate from structural imperfections at
the interfaces and in the bulk. We find that the magnetization
relaxation is logarithmic. Here the magnetic viscosity
is found to increase first with increasing temperature, then it reaches
a maximum around T$_g$, and then it decreases with increasing temperature.
This behavior is different from that of conventional spin glasses
where the logarithmic creep rate is observed to increase with
temperature. Power law also gives good fits and it is better than the logarithmic
fit at higher temperatures. The dynamical
effects of these multilayers are related to the 
relaxation of thermally blocked superparamagnetic grains and 
magnetic domains in the film layers.

\end{abstract}

                             

\maketitle

\section{INTRODUCTION}
In Fe-Cr GMR multilayers the ferromagnetic (FM)
Fe layers are exchange coupled through the non-magnetic 
Cr spacer layers. An antiferromagnetic arrangement of the Fe layers is 
engineered by varying Cr spacer layer thickness. With varying Cr 
thickness successive Fe layers show oscillatory antiferromagnetic 
(AFM) and FM couplings but with decreasing peak coupling strength.
It is believed that in an external magnetic field $H>H_{sat}$, the magnetization
in the Fe layers are aligned as in a ferromagnetic material but in 
the absence of the magnetic field they are in antiferromagnetic configuration.
Our study is focussed at finding the magnetic relaxation of these multilayers
in low external magnetic fields ($\sim$100 Oe).

Thermoremanent magnetization is a thermally activated process. When the 
applied magnetic field is removed the magnetization tries to approach 
the remanent magnetization value in order to minimise the energy of 
the system. A magnetic system has dipolar energy, anisotropy energy, 
and exchange energy. In general, ferromagnetic materials do not 
respond to the magnetic fields immediately. There is a time lag between 
the application/withdrawl of the magnetic field and the response of magnetization 
to the field. This phenomenon is called ``magnetic viscosity''
\cite{Street:1949}. Folks and Street \cite{Folks:1994} described
this time lag by domain processes which progress through states of metastable
equilibrium to a stable state. The major domain processes are:
\begin{enumerate}
\item coherent or incoherent rotation of magnetization in single domain particles,
\item pinning and unpinning of domain boundary walls, and
\item nucleation of domains of reverse magnetization.
\end{enumerate}

Coherent rotation of magnetization vector and the
Bloch-wall formation are primary consequences of lowering energy after 
the external field is applied/removed. Thermal agitation plays
an important role in transition from metastable to stable
states. One may ask what are the time scales to achieve stable 
states for paramagnetic, FM, and AFM materials. What is the role of magnetic
interaction and crystal structure in the time dependence of magnetization?

Dahlberg et al. \cite{Dahlberg:1994} explained the time 
dependence as a consequence of interactions or couplings. The 
interaction between relaxing spins, the dipole coupling, drives the
system from an initial state towards the equilibrium state. They attributed the 
strong time dependence of the magnetization in Co-Cr films to the 
demagnetization in the film.

Sinha \cite{Sinha:1996} et al. found that in $Fe_{80-x}Ni_xCr_{20}$ ($x$ = 30,
FM phase) M(t) fits well to the power law decay. They found that with the increase
of wait time, the exponent becomes smaller. For $x$ =14 (AF phase) also,
power law decay describes M(t) quite well and the 
exponent decreases with increasing temperature.
Chamberlin \cite{Chamberlin:1991} studied EuS which is considered to be an
ideal Heisenberg FM system with a Curie temperature $\sim 16.6$ K.
They found that the plots of remanent magnetization
versus logarithm of time show negative curvature for $T<T_B$(=17.75 K) and 
exhibits an S-shaped curve for $T>T_B$ . Ulrich et al.
\cite{Ulrich:2003} studied the magnetic relaxation of single
domain ferromagnetic particles below the blocking temperature. 
They found that for all particle densities the relaxation decays following
a power law, with density-dependent exponent and a temperature-dependent
prefactor. They used Monte Carlo simulations to study the 
influence of dipolar interactions and polydispersion on the magnetic 
relaxation of single-domain FM particles below the blocking temperature.
They concluded (i) stretched exponential decay at low densities, (ii) a power
law decay at intermediate densities, and (iii) relaxation toward a non-
vanishing remanent magnetization at high densities.

In an earlier study of $Cu_{100-x}Mn_x(x$ = 76, 83 and both in AF phase) we found 
\cite{Patel:2002} that the power law
decay is better at lower temperatures. The fits at a given temperature 
improve with stronger long-range antiferromagnetic order (AF1 structure).
Leighton and Schuller \cite{Leighton:2001} used time-dependent magnetization
measurements to probe the asymmetry in the magnetization reversal mechanism
in exchange-biased $MnF_2/Fe$ bilayers. They found that on one side of the loop,
coherent rotation of magnetization plays an important role while on the other side
the domain nucleation. The time dependence of the magnetization on the
coherent rotation side of the loop had a form that is consistent with a small 
distribution of energy barriers. The side of the loop characterised by domain
nucleation and propagation shows a logarithmic time dependence with a field
dependent viscosity. Similar investigation was also carried by
Fullerton and Bader \cite{Fullerton:1996} on Fe/FeSi multilayers.

Chamberlin et al. \cite{Chamberlin:1984} studied the time decay of the
thermoremanent magnetization in 1.0 \% CuMn and 2.6 \% AgMn spin glasses.
They found it to be a stretched exponential function. Till that time no data had been
published supporting an algebraic decay (power law) for magnetization.
Chamberlin also \cite{Chamberlin:1984b} found that the effect of wait-
time $t_w$ can be empirically characterised as an exponential decrease
of the relaxation frequency with increasing wait time. If the sample is 
allowed to remain in the field-cooled state long enough before the field 
is removed, then the magnetization will not relax: a spin glass can have 
a permanent magnetization in zero field. Chubykalo and Gonzalez
\cite{Chubykalo:1999} simulated the relaxational behavior of
Co-Ni multilayers with different layer thickness. Their simulations
had shown that the thermally activated demagnetization process in 
Co-Ni multilayers does not occur according to the Arrhenius kinetics.
Panagiotopoulos et al. \cite{Panagioto:2002} found that the
magnetic relaxation follows the ln(t) behavior at 5 K in La-Ca-Mn-O
FM/AF multilayer with $T_B$ = 70 K. In FM films that exhibit a wide
range of energy barriers the magnetization time decay follows a ln(t)
behavior only below a blocking or freezing temperature, resulting from
the superposition of many exponential decays with different magnetic
relaxation times. Chen et al. \cite{Chen:2003} found that the magnetization 
relaxation of a $[Co_{80}Fe_{20}(1.4 nm)/Al_2O_3(3 nm)]_{10}$ sample
obeys a power-law decay of the thermoremanent magnetic moment with
an asymptotic remanence when starting from a completely relaxed FC state.

Street and Brown \cite{Street:1994} described two types of mechanisms that are
responsible for the time dependence in ferromagnetic materials known as 
``diffusion'' and ``fluctuation'' after-effect or viscosity. Chantrell et al.
\cite{Chantrell:1994} gave a phenomenological theory based on the intrinsic
energy barrier to explain the form of time dependence of the magnetization.
The slow relaxation is related to the irreversible magnetic behavior via a 
fictitious fluctuation field $H_f$ which itself determines a quantity called
the activation volume $V_{act}$. Both $H_f$ and $V_{act}$ are related to
the magnetization reversal process. For granular materials $V_{act}$ is generally
smaller than the grain size. A simple model for the time dependence of the switching 
field in magnetic recording media by Sharrock \cite{Sharrock:1994} accounts for thermally
assisted crossing of an energy barrier whose height is reduced by the applied field. Goodman
et al. \cite{Goodman:2000} studied a system with a narrow distribution of energy barriers.
In such systems, at fields less than the coercive field, an accelerating variation of magnetization
with log of time is seen and a decelerating behavior of field above the coercivity.
In the coercive field region an `S' shaped variation is observed.

\section{EXPERIMENTAL DETAILS}
Our samples were grown on Si substrates by ion beam sputter deposition
technique using Xe ion at 900 V with beam current of 20 mA and 
1100 V with beam current of 30 mA. 
The typical structures are Si/Cr(50 \AA)/[Fe(20 \AA)/Cr(t \AA)]$\times$
30/Cr(50 -t \AA). Samples 1 - 4 have t = 6, 8, 10 and 12 \AA, respectively. 
Samples 1 and 2 are deposited at 900 V and samples 3 and 4 are at 
1100 V. Our multilayers show a GMR $(=((\rho(H,T)-\rho(0,T))/\rho(0,T))
\times 100 \%)$ of $\approx$ 1, 33, 20 and 21 \% at 4.2 K for an 
external field of $\approx$ 1 tesla ($H_{sat}$). These samples are 
well characterised and the details have been 
given elsewhere\cite{Lannon:2002}. 
All the experiments were done with a Quantum Design superconducting 
quantum interference device (SQUID) magnetometer(MPMS). A magnetic 
field of 100 Oe is applied in the plane of the multilayer samples
at 300 K and the sample is cooled down to the measuring temperature. After
the temperature is stabilised, the magnetic field is set to zero and
M(T) measurements were started and continued till 12,000 s.

\section{THEORY}
Chamberlin \cite{Chamberlin:1994} summarised the time dependence 
of magnetization in terms of different mathematical expressions as follows:

\begin{enumerate}
\item The most popular empirical expression for characterising amorphous
materials has been the Kohlrausch-Williams-Watts stretched exponential
$M(t) \propto exp(-(t/\tau)^\beta) $.
\item For crystals, the Curie-von Schweidler power law $M(t) \propto 
t^{-\beta}$.
\item For magnetic aftereffects, the Ne\`{e}l logarithmic time 
dependence $M(t) \propto log(t/\tau)$ is popular.
\end{enumerate}
When the applied magnetic field is removed, the magnetization takes 
finite time to cross the energy barrier for reversal. If the energy 
barriers are identical then the magnetic moment
$M$ is given by \cite{Leighton:2001}

\begin{equation}
\label{eq:exp}
M(t) = A + B \ exp(-t/\tau_0),
\end{equation}

\noindent
where

\begin{equation}
\frac{1}{\tau_0} = f_0 \ exp\left(\frac{E_A}{k_BT}\right),  \nonumber
\end{equation}
$A$ and $B$ are constants, $t$ represents
the elapsed time, $f_0$ is the attempt frequency, $E_A$ is the
activation energy, and T is the 
measurement temperature. When there is a distribution of barriers, then

\begin{equation}
\label{eq:ln}
M(t) = M_0 - S \ ln(t).
\end{equation}

Here S is called the ``magnetic viscosity" and $M_0$ is a constant 
at a given measuring field. S is expected to reach a peak value near
the coercive field $H_c$ where the rate of change of the moment with 
time reaches the maximum. The intermediate case with a small 
distribution of barrier heights is more complicated to quantify.

Power-law behavior is predicted by scaling theories for domain growth 
and internal dynamics \cite{Chamberlin:1991}. Calculations based on
Sherrington-Kirkpatrick mean-field model suggest that the magnetization
decays algebraically as

\begin{equation}
\label{eq:power}
M(t) = M_0 \ t^{-\beta}.
\end{equation}

Other treatments yield Kolrausch-Williams-Watts stretched-exponential
relaxation which is common in spin-glass like systems 
\cite{Chamberlin:1994} given by

\begin{equation}
\label{eq:strexp}
M(t) = M_0 \ exp\left( -\left( \frac{t}{\tau} \right)^n \right).
\end{equation}

When measured in a
small static field, the temperature dependence of the magnetization of a
spin glass changes abruptly at the glass transition temperature ($T_g$).
Above $T_g$ the magnetization obeys the Curie-Weiss law, attributable 
to weakly interacting paramagnetic spins whereas below $T_g$ the 
magnetization is cooling history dependent. If it is field cooled then the
magnetization is nearly independent of temperature below $T_g$; if it is zero-field 
cooled then the magnetization increases with increasing temperature.

\section{RESULTS AND DISCUSSION}
\begin{figure}
\includegraphics[bb=1pt 13pt 393pt 195pt, width=8.5cm]{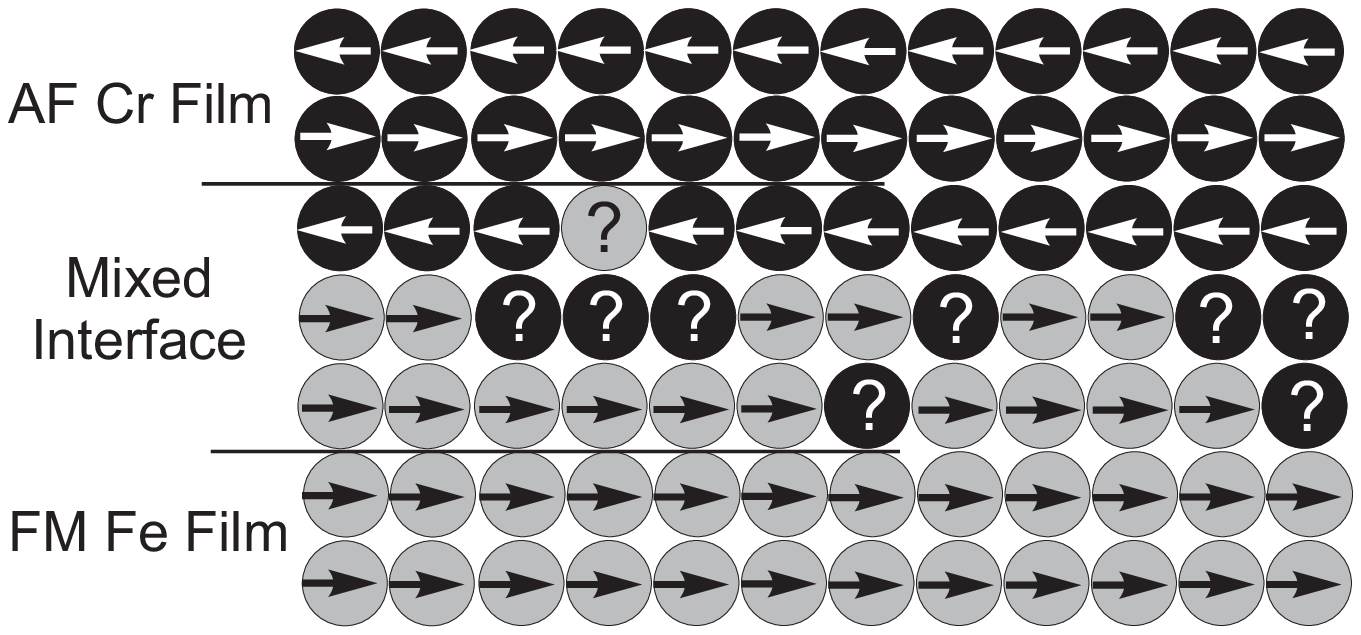}
\caption{\label{fig:sample} Schematic diagram of the sputtered GMR samples. Black 
and grey spheres represent Cr and Fe atoms, respectively. Typical 
estimates of the relative strengths of the Fe-Fe, Fe-Cr, and Cr-Cr 
couplings are 1 : -0.3 : -0.18 \cite{Shi:1997} and 1 : -0.55 : 
-0.3 \cite{Berger:1997}. Directions of the arrow show the spin alignment
direction. `?' shows the uncertainty(frustration) in the spin direction at that site.}
\end{figure}

Structure-wise our samples are a good combination of crystalline 
layers of ferromagnetic Fe and antiferromagnetic Cr and the polycrystalline interfaces.
So these samples provide a complicated combination of crystalline layers 
and polycrystalline interfaces consisting of ferromagnetic, 
antiferromagnetic and spin-glass like structures as shown schematically
in Fig. \ref{fig:sample}.  In this figure black and grey spheres represent Cr and Fe atoms,
respectively. Directions of the arrow show the direction of spin alignment.
Typical estimates of the relative strengths of the Fe-Fe, Fe-Cr, and Cr-Cr 
couplings are 1 : -0.3 : -0.18 \cite{Shi:1997} and 1 : -0.55 : 
-0.3 \cite{Berger:1997}. In other words, Fe-Fe coupling is FM and Fe-Cr
\& Cr-Cr couplings are AF. Fe-Fe coupling is two/three times stronger than Fe-Cr coupling.
`?' in Fig. \ref{fig:sample} represents uncertainty of spin direction(frustration) at that site. For a particular direction,
either it will be violating the FM Fe-Fe coupling or the AF Fe-Cr coupling. As the Fe-Fe coupling
is stronger than the Fe-Cr and Cr-Cr coupling most of the frustration occurs at Cr sites (black spheres).
Attempts to explain magnetic relaxation with mathematical 
expressions began a century back. We have made a similar attempt to understand
the response of the combined system for small magnetic fields. Although
Chamberlin\cite{Chamberlin:1994} argued that these empirical expressions
are simple mathematical formulae that give generally good agreement with
a wide variety of measurements but demonstrating agreement with these 
formulae tells nothing about the mechanism of response.

For our analysis, we have used a standard non-linear least-squares-fit
program. Here $\chi^2$ is defined as
\begin{equation}
\chi^2 = \frac{1}{N} \sum_{i=1}^{N} 
\frac{(M_{i-measured}-M_{i-fitted})^2}{M_{i-mean}^2}.
\end{equation}

In our multilayer
system, the ferromagnetic Fe layers have in-plane magnetization but as 
they are not super-lattices, there will be some domain 
like structures in the ferromagnetic plane with distribution of
domain volume. Thus we expect that there is a
distribution of barriers for which logarithmic relaxation of magnetization is generally found.
So we first tried the logarithmic fit in the form of 
Eq. (~\ref{eq:ln}). For our analysis we have taken data from 100 s to
12,000 s. We have made measurements with a SQUID magnetometer and it 
takes $\sim$ 10 s to make one measurement and a much longer time to 
achieve stability and equilibrium condition after removal of the 
magnetic field. So we have not analysed the data for t $<$ 100 s.

\begin{figure}
\includegraphics[bb=0.4cm 0.5cm 9.0cm 7.7cm,width=8.5cm]{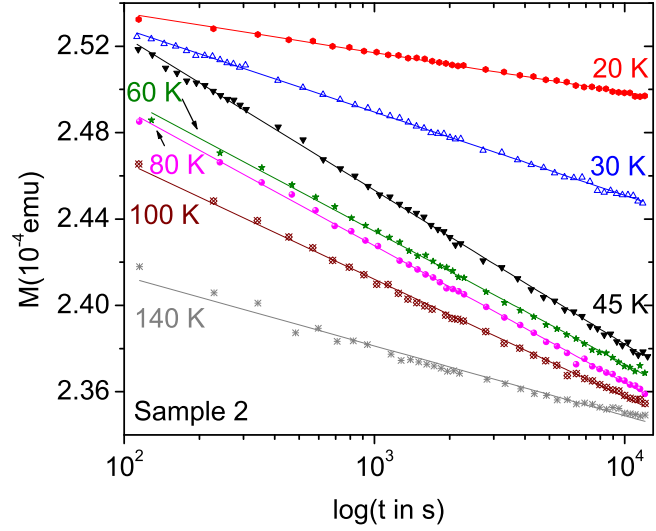}
\caption{\label{fig:S2ln} (Color online) M (raw data) is 
plotted against {\it log} (t in s) for
sample 2. The solid lines are the logarithmic fits to Eq. (~\ref{eq:ln}).
The data and the fits are multiplied by 1.285, 1.66, 1.73, 2.27, 2.85, and 3.92
for 30, 45, 60, 80, 100, and 140 K, respectively to show all the
curves on the same figure.}
\end{figure}

\begin{figure}
\includegraphics[bb=0.4cm 0.5cm 9.0cm 7.7cm,width=8.5cm]{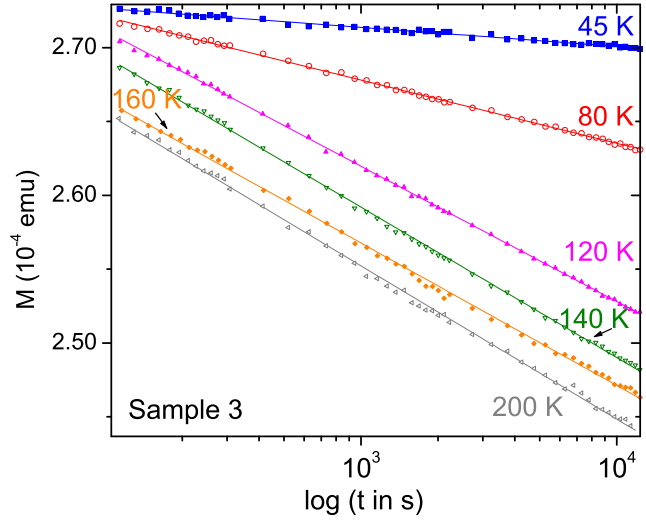}
\caption{\label{fig:x4alllog} (Color online) M (raw data) is 
plotted against {\it log} (t in s) for
sample 3. The solid lines are the logarithmic fits to Eq. (~\ref{eq:ln}).
The data and the fits are multiplied by 1.3, 2.3, 3.2, 4.42, and 8.4
for 80, 120, 140, 160, and 200 K, respectively to show all the
curves on the same figure.}
\end{figure}

\begin{figure}
\includegraphics[bb=0.4cm 0.5cm 9.0cm 7.7cm,width=8.5cm]{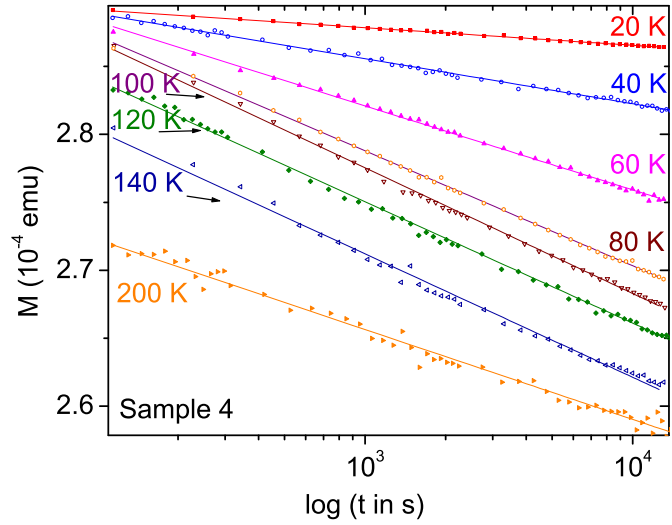}
\caption{\label{fig:x6alllog} (Color online) M (raw data) is 
plotted against {\it log} (t in s) for
sample 4. The solid lines are the logarithmic fits to Eq. (~\ref{eq:ln}).
The data and the fits are multiplied by 1.19, 1.45, 1.9, 2.5, 3.45, 4, and 8.2
for 40, 60, 80, 100, 120, 140, and 200 K, respectively 
to show all the curves on the same figure.}
\end{figure}

\begin{figure}
\includegraphics[bb=14pt 16pt 263pt 222pt,width=8.5cm]{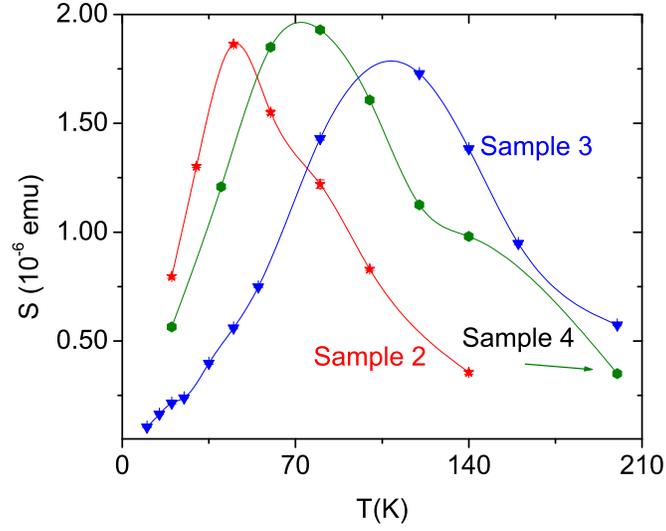}
\caption{\label{fig:pwrp} (Color online) Variation of the parameter S of  Eq. 
(~\ref{eq:ln}) with temperature. The solid lines are just guides 
to the eye.}
\end{figure}

\begin{figure}
\includegraphics[bb=0.4cm 0.5cm 9.0cm 7.7cm,width=8.5cm]{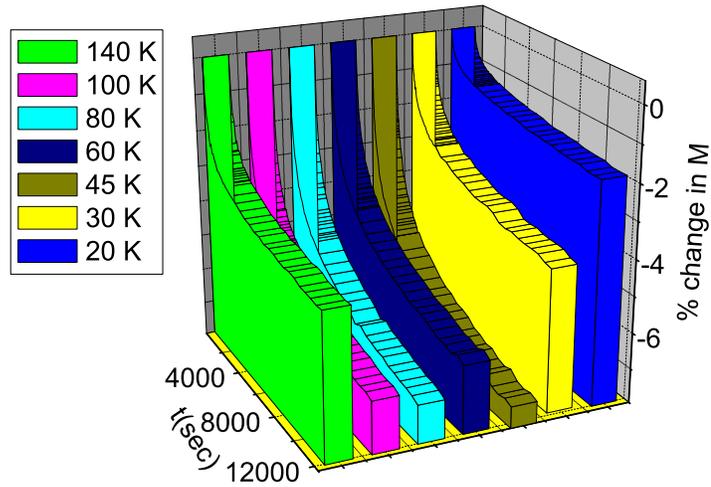}
\caption{\label{fig:f83d} (Color online) Percentage change in 
M is plotted against t (s) at various temperatures for
sample 2.}
\end{figure}

Logarithmic relaxation gives excellent fits for samples 2 - 4 with correlation coefficients
$\sim$ 0.997 and the normalised $\chi^2$ consistent with the experimental 
resolution of one part in $10^4$. For sample 1,
which has the least GMR, the fit is not so good ($R^2 \sim
0.90$). Figures \ref{fig:S2ln}, \ref{fig:x4alllog}, and \ref{fig:x6alllog}
show the raw M vs. ln t data and the fits for samples 2, 3, and 4,
respectively at different temperatures. For each sample,
$M_0$ decreases with increasing temperature which can
be understood as the decrease in magnetization due to thermal 
fluctuations. The coefficient S, which is termed as the magnetic viscosity, 
initially increases linearly with temperature; then it decreases at higher
temperatures. Larger S implies larger change in the magnetization during the
observational time period. S shows a peak around $T_g$ of the ZFC 
magnetization curves (samples 2 - 4 show $T_g$ at $\sim$ 62, 123, and 83 K,
respectively taken at 100 Oe \cite{Patel:2005}).
This can be understood as follows: At temperatures much lower than 
the freezing temperature, due to the low temperature (less thermal 
energy), the magnetic moments can not relax much (as they are nearly frozen). So 
lower the temperature lower is the value of S. At temperatures
much higher than the freezing temperature, due to higher thermal energy 
the system behaves more and more like a paramagnetic/superparamagnetic 
system. In the paramagnetic region the entire magnetization relaxes very
rapidly ($<$10 s)\cite{Chamberlin:1984}. So in our measurement time window
($10^2 - 10^4$ s) we hardly observe any relaxation and hence S looks small.
At around $T_g$ in the ZFC magnetization curve, the
thermal energy provides the moments freedom to move and they can also 
interact with one another. In this region, the system is like a spin glass
which is about to unfreeze and so the
magnetization responds much faster. That is why we find maxima in 
the magnetic viscosity vs. temperature curves. This is shown in 
Fig. \ref{fig:pwrp}. To make the comparison more clear we have shown the percentage
change in magnetization with time at different temperatures for sample 2 in Fig. \ref{fig:f83d}.
Similar variation of S with temperature was found by Guy \cite{Guy:1978} in a
Au-Fe spin-glass alloy containing 2 at. \% Au. However, Lottis et al. \cite {Lottis:1988} found 
broad maxima in the temperature dependence of S in Co-Cr films. 

\begin{figure}
\includegraphics[bb=0.4cm 0.5cm 9.0cm 7.7cm,width=8.5cm]{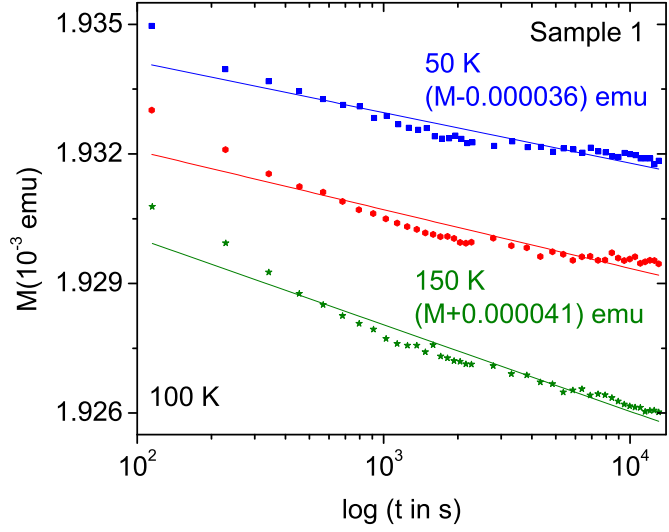}
\caption{\label{fig:S1ln} (Color online) M(raw data) is 
plotted against {\it log} (t in s) for
sample 1. The solid lines are the logarithmic fits to Eq. (~\ref{eq:ln}). 
The data and the fits for 50 and 150 K are shifted along y-axis for clarity.}
\end{figure}

Figure  \ref{fig:S1ln} shows the raw M vs. ln t data and the fit for sample 1. Sample 1
showed the least GMR. The interpretation for the least GMR is that in this
sample with Cr thickness of 6 \AA \ most Fe layers are coupled ferromagnetically to the neighbouring Fe
layers. So the whole sample is like a FM material. The ZFC magnetization measurements
of this sample show that the magnetization decreases monotonically with 
increasing temperature. So measurement temperatures of 50,
100, and 150 K are much above the $T_B$ of this sample. The magnetization,
plotted on a logarithmic scale, has clearly an `S'- shape as shown in Fig. \ref{fig:S1ln}.

\begin{figure}
\includegraphics[bb=0.4cm 0.5cm 9.0cm 7.7cm,width=8.5cm]{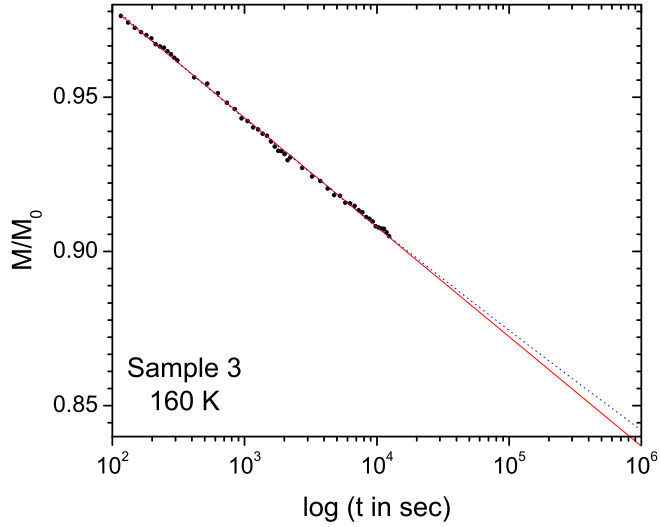}
\caption{\label{fig:S3Ex} (Color online) M/M$_0$ is plotted against {\it log} (t in 
s) for sample 3 at 160 K. The solid line is the logarithmic 
fit to Eq. (~\ref{eq:ln}) and the dashed line is the power fit to Eq. (~\ref{eq:power}).
Both the fits are extended till $10^6$ s to show that in larger time window 
the power fit decreases slower than the logarithmic fit. However, in our
measurement time span there is only a small difference between the two fits.}
\end{figure}

Next we tried to fit the data to the power law in the form of Eq. 
(~\ref{eq:power}). We find that the power law
gives equally good fits for samples 2 - 4, with $R^2 >$ 0.99 along with
$\chi^2$ consistent with the experimental resolution. 
The numerical value of the parameter $\beta$
is the same as of the parameter `S' divided by $M_0$ of Eq. (~\ref{eq:ln}). We can rewrite
Eq. (~\ref{eq:power}) as

\begin{equation}
\label{eq:power1}
M(t) = M_0 \ t^{-\beta} = M_0(e^{-\beta \ ln(t)}) \\
\approx M_0 - (M_0\beta) \ ln(t),
\end{equation}
since $\beta$ is very small and so we can neglect higher order terms. This has
a form similar to that of Eq. (~\ref{eq:ln}). For larger time scale $t > 10^5$ s, $t^{-\beta}$
decreases faster than $ln(t)$. We have tried to show this feature in Fig. 
\ref{fig:S3Ex}. In other words, from our measurement time window we shall not
be able to distinguish log and power fits. However, a careful observation from Tables I-IV 
indicates a small difference between the two fits. Eq.
(~\ref{eq:ln}) gives better fit than Eq. (~\ref{eq:power}) at low temperatures
but at higher temperatures (much above $T_g$) Eq. (~\ref{eq:power}) gives better
fits. Sample 1 does not have $T_g$ or $T_B$ in the measured temperature
range and so the power law fit is better for all the temperatures. A stronger proof would 
have been possible if we could take data till $10^5 - 10^6$ s.

Next we tried the stretched exponential fit in the form of Eq. (~\ref{eq:strexp}).
The low-field magnetization measurements on samples 2-4 had shown
spin-glass-like history dependent behaviour. But this fitting gave unrealistic
large time constants $(\tau)$ with equally large errors.

To conclude, in these sputtered GMR multilayers which are structural combinations
of FM Fe films, AF Cr Films, and Fe-Cr mixed interfaces, the magnetization decays
logarithmically at low temperatures and above $T_g$ the magnetization decays
algebraically. In the multilayers one expects less distribution of energy barriers
but we could not find any exponential decay. For `AFM' (sample 2, 3, 
and 4) and `FM' (sample 1) multilayers we found different decay mechanisms. The power law decay
is better at higher temperatures. It is difficult to distinguish between `good
AFM' (sample 2 with GMR of 33 \%) and `AFM' (sample 3 with GMR of 21 \%
and sample 4 with GMR of 20 \%) samples from their decay behavior.

\begin{acknowledgments}
We sincerely thank Drs. D. Temple and C. Pace of MCNC, Electronic Technologies
Division, Research Triangle Park, North Carolina for providing us the samples.
One of us (R.S.P.) acknowledges CSIR, Government of India, for financial support.
\end{acknowledgments}

\vspace{15cm}

\begin{table}
\caption{Values of $\chi^2$, correlation coefficient 
R$^2$, the parameters $M_0$, $S$ of Eq. (~\ref{eq:ln}), $M_0$, and
$\beta$ of Eq. (~\ref{eq:power}) for sample 1.}
\begin{tabular}{|c|c|c|c|c|} \hline
T(K) & $\chi^{2}(10^{-8})$ & R$^{2}$ & M$_0(10^{-5} emu)$  & S(10$^{-7} emu)$ \\ \hline
50 &  1.407 & 0.8778 & 197.24 $\pm $ 0.02 & 5.1 $\pm $ 0.3 \\ \hline
100 & 2.110 & 0.8707 & 193.48 $\pm $ 0.03 & 5.9 $\pm $ 0.3 \\ \hline
150 & 1.713 & 0.9497 & 189.30 $\pm $ 0.02 & 8.7 $\pm $ 0.3 \\ \hline \hline
T(K) & $\chi^{2}(10^{-8})$ & R$^{2}$ & M$_0(10^{-5} emu)$ & $\beta$(10$^{-4})$ \\ \hline
50 & 1.405 & 0.8779 & 197.00 $\pm $ 0.02 & 2.6 $\pm $ 0.2 \\ \hline
100 & 2.108 & 0.8708 & 193.00 $\pm $ 0.03 & 3.1 $\pm $ 0.2 \\ \hline
150 & 1.709 & 0.9498 & 189.00 $\pm $ 0.02 & 4.6 $\pm $ 0.2 \\ \hline
\end{tabular}
\label{tab:F06}
\end{table}

\begin{table}
\caption{Values of $\chi^2$, correlation coefficient R$^2$,
and the parameters $M_0$, $S$ of Eq. (~\ref{eq:ln}), $M_0$, and
$\beta$ of Eq. (~\ref{eq:power}) for sample 2.}
\begin{tabular}{|c|c|c|c|c|}\hline
T(K) & $\chi^{2}(10^{-8})$ & R$^{2}$ & M$_0(10^{-5} emu)$ & S(10$^{-7} emu)$ \\ \hline
20 & 5.1674 & 0.9965 & 25.721 $\pm $0.006 & 7.96 $\pm $ 0.08 \\ \hline
30 & 11.621 & 0.9988 & 20.274 $\pm $ 0.004 & 13.03 $\pm $ 0.06 \\ \hline
45 & 29.432 & 0.9992 & 16.066 $\pm $ 0.006 & 18.64 $\pm $ 0.07 \\ \hline
60 & 25.028 & 0.9986 & 15.140 $\pm $ 0.008 & 15.5 $\pm $ 0.1 \\ \hline
80 & 20.487 & 0.9895 & 11.54 $\pm $ 0.01 & 12.2 $\pm $ 0.2 \\ \hline
100 & 24.683 & 0.9983 & 9.037 $\pm $ 0.004 & 8.30 $\pm $ 0.05 \\ \hline
140 & 89.531 & 0.9833 & 6.320 $\pm $ 0.006 & 3.56 $\pm $ 0.08 \\ \hline
200 & 1119.8 & 0.9969 & 1.698 $\pm $ 0.004 & 6.21 $\pm $ 0.06 \\ \hline \hline
T(K) & $\chi^{2}(10^{-8})$ & R$^{2}$ & M$_0(10^{-5} emu)$ & $\beta$(10$^{-3})$ \\ \hline
20 & 5.4684 & 0.9963 & 25.727 $\pm $0.006 & 3.17 $\pm $ 0.03 \\ \hline
30 & 12.7867 & 0.9987 & 20.296 $\pm $ 0.005 & 6.74 $\pm $ 0.03 \\ \hline
45 & 36.7689 & 0.999 & 16.126 $\pm $ 0.007 & 12.65 $\pm $ 0.05 \\ \hline
60 & 29.2681 & 0.9983 & 15.187 $\pm $ 0.009 & 11.06 $\pm $ 0.07 \\ \hline
80 & 22.0195 & 0.9988 & 11.570 $\pm $ 0.006 & 11.41 $\pm $ 0.06 \\ \hline
100 & 19.4450 & 0.9987 & 9.059 $\pm $ 0.004 & 9.86 $\pm $ 0.06 \\ \hline
140 & 88.8819 & 0.9834 & 6.328 $\pm $ 0.006 & 5.9 $\pm $ 0.1 \\ \hline
200 & 1807.83 & 0.995 & 1.7690 $\pm $ 0.007 & 48.3 $\pm $ 0.6 \\ \hline
\end{tabular}
\label{tab:F08}
\end{table}

\begin{table}[htbp]
\caption{Values of $\chi^2$, correlation coefficient R$^2$,
the parameters $M_0$, $S$ of Eq. (~\ref{eq:ln}),  $M_0$, and $\beta$
Eq. (~\ref{eq:power}) for sample 3.}
\begin{tabular} {|c|c|c|c|c|} \hline
T(K) & $\chi^{2}(10^{-8})$ & R$^{2}$ & M$_0(10^{-5} emu)$ & S(10$^{-7} emu)$ \\ \hline
10  & 15.52408 & 0.9872 & 26.223 $\pm $ 0.002 & 1.04 $\pm $ 0.02 \\ \hline
15  & 17.2138 & 0.9944 & 25.8571 $\pm $ 0.002 & 1.64  $\pm $ 0.02 \\ \hline
20 & 63.2449 & 0.9885 & 25.5907 $\pm $ 0.003 & 2.15 $\pm $ 0.04 \\ \hline
25 & 44.2073 & 0.9930 & 26.2994 $\pm $ 0.003 & 2.39 $\pm $ 0.03 \\ \hline
35 & 87.8776 & 0.9954 & 34.2492 $\pm $ 0.004 & 3.97 $\pm $ 0.04 \\ \hline
45  & 12.3645 & 0.9872 & 27.525 $\pm $ 0.007 & 5.60 $\pm $ 0.09 \\ \hline
55  & 54.5099 & 0.9993 & 22.8914 $\pm $ 0.002 & 7.47 $\pm $ 0.03 \\ \hline
80  & 12.2490 & 0.9989 & 21.588 $\pm $ 0.005 & 14.3  $\pm $ 0.07 \\ \hline
120 & 16.2190 & 0.9997 & 12.584 $\pm $ 0.003 & 17.29 $\pm $ 0.04 \\ \hline
140 & 36.9160 & 0.9994 & 9.056 $\pm $ 0.003 & 13.83 $\pm $ 0.05 \\ \hline
160 & 75.9658 & 0.9988 & 6.464 $\pm $ 0.003 & 9.48  $\pm $ 0.05 \\ \hline
200 & 84.3582 & 0.9988 & 3.409 $\pm $ 0.002 & 5.37 $\pm $ 0.03 \\ \hline \hline
T(K) & $\chi^{2}(10^{-8})$ & R$^{2}$ & M$_0(10^{-5} emu)$ & $\beta$(10$^{-3})$ \\ \hline
10 & 0.3085 & 0.9872 & 29.814 $\pm $ 0.002 & 0.397 $\pm $ 0.004 \\ \hline
15 & 0.3483 & 0.9944 & 29.398 $\pm $ 0.002 & 0.635  $\pm $ 0.008 \\ \hline
20 & 1.2901 & 0.9884 & 29.096 $\pm $ 0.003 & 0.85 $\pm $ 0.02 \\ \hline
25 & 0.8924 & 0.9930 & 29.901 $\pm $ 0.003 & 0.92 $\pm $ 0.02 \\ \hline
35 & 1.8190 & 0.9953 & 28.847 $\pm $ 0.004 & 1.58  $\pm $ 0.02 \\ \hline
45 & 12.3258 & 0.9872 & 27.527 $\pm $ 0.007 & 2.06 $\pm $ 0.03 \\ \hline
55 & 1.1379 & 0.9993 & 26.034 $\pm $ 0.003 & 3.35 $\pm $ 0.02 \\ \hline
80 & 13.4735 & 0.9988 & 21.613 $\pm $ 0.005 & 6.95  $\pm $ 0.03 \\ \hline
120 & 22.8617 & 0.9996 & 12.652 $\pm $ 0.004 & 15.24 $\pm $ 0.04 \\ \hline
140 & 26.3087 & 0.9996 & 9.116 $\pm $ 0.003 & 17.15 $\pm $ 0.05 \\ \hline
160 & 56.4284 & 0.9991 & 6.504 $\pm $ 0.003 & 16.39  $\pm $ 0.07 \\ \hline
200 & 5296.38 & 0.9456 & 3.364 $\pm $ 0.001 & 15.0 $\pm $ 0.4 \\ \hline
\end{tabular}
\label{tab:X4}
\end{table}

\begin{table}[htbp]
\caption{Values of $\chi^2$, correlation coefficient R$^2$,
and the parameters $M_0$, S of Eq. (~\ref{eq:ln}), 
$M_0$, and $\beta$ of Eq. (~\ref{eq:ln}) for sample 4.}
\begin{tabular} {|c|c|c|c|c|} \hline \hline
T(K) & $\chi^{2}(10^{-8})$ & R$^{2}$ & M$_0(10^{-5} emu)$ & S(10$^{-7} emu)$ \\  \hline
20 & 1.8586 & 0.9969 & 29.175 $\pm $ 0.004 & 5.64 $\pm $ 0.05 \\ \hline
40 & 20.7856 & 0.9969 & 24.832 $\pm $ 0.007 & 12.08 $\pm $ 0.09 \\ \hline
60 & 27.4387 & 0.9980 & 20.73 $\pm $ 0.01& 18.5 $\pm $ 0.1 \\ \hline
80 & 28.9706 & 0.9989 & 16.007 $\pm $ 0.008 & 19.3 $\pm $ 0.1 \\ \hline
100 & 49.0388 & 0.9985 & 12.212 $\pm $ 0.008 & 16.07 $\pm $ 0.1 \\ \hline
120 & 103.588 & 0.9981 & 8.749 $\pm $ 0.005 & 11.25 $\pm $ 0.05 \\ \hline
140 & 248.365 & 0.9925 & 7.46 $\pm $ 0.01 & 9.8 $\pm $ 0.1 \\ \hline
200 & 346.956 & 0.9883 & 3.481 $\pm $ 0.004 & 3.50 $\pm $ 0.05 \\ \hline \hline
T(K) & $\chi^{2}(10^{-8})$ & R$^{2}$ & M$_0(10^{-5} emu)$ & $\beta$(10$^{-3})$ \\ \hline
20& 26.4418 & 0.9955& 2.9180 $\pm $ 0.0005& 1.97 $\pm $ 0.02 \\ \hline
40& 20.4407 & 0.9969& 2.4848 $\pm $ 0.0007& 5.05 $\pm $ 0.04 \\ \hline
60& 33.6017 & 0.9976& 2.078 $\pm $ 0.001& 9.58 $\pm $ 0.07 \\ \hline
80& 33.0043 & 0.9988& 1.608 $\pm $ 0.001& 13.28 $\pm $ 0.07 \\ \hline
100& 32.6814 & 0.9990& 1.2280 $\pm $ 0.0007& 14.62 $\pm $ 0.07 \\ \hline
120& 79.7989 & 0.9985& 0.8793 $\pm $ 0.0005& 14.22 $\pm $ 0.07 \\ \hline
140& 194.536 & 0.9942& 0.750 $\pm $ 0.001& 14.7 $\pm $ 0.2 \\ \hline
200& 333.72 & 0.9887& 0.3492 $\pm $ 0.0004& 10.9 $\pm $ 0.1 \\ \hline
\end{tabular}
\label{tab6:X9004pwr1}
\end{table}

\end{document}